\documentclass[preprint,12pt]{elsarticle}
\usepackage{amssymb}
\journal{Physics Letters B}
\def\be{\begin{equation}}
\def\ee{\end{equation}}
\def\bea{\begin{eqnarray}}
\def\eea{\end{eqnarray}}

\def\L{{\cal{L}}}
\begin{document}
\begin{frontmatter}
\title{\bf Braneworld setup and embedding in teleparallel gravity }
\author{A. Behboodi}%
\ead{a.behboodi@stu.umz.ac.ir} \address{Department of Physics,
Faculty of Basic Sciences,
University of Mazandaran, P. O. Box 47416-95447, Babolsar, IRAN,}%
\author{S. Akhshabi}%
\ead{s.akhshabi@gu.ac.ir} \address{ Department of Physics, Golestan
University, P. O. Box 49138-15759, Gorgan, IRAN,}
\author{K. Nozari}%
\ead{knozari@umz.ac.ir} \address{Department of Physics, Faculty of
Basic Sciences,
University of Mazandaran, P. O. Box 47416-95447, Babolsar, IRAN}%

\vspace{2cm}
\begin{abstract}
We construct the setup of a five-dimensional braneworld scenario in
teleparallel gravity. Both cases of  Minkowski and
Friedmann-Robertson-Walker branes embedded in Anti de Sitter bulk
are studied and the effective 4-D action were studied. 4-dimensional
local Lorentz invariance is found to be recovered in both cases.
However, due to different junction conditions, the equations
governing the 4-D cosmological evolution differ from general
relativistic case. Using the results of Ref. \cite{BAN}, we consider
a simple inflationary scenario in this setup. The inflation
parameters are found to be modified compared to general relativistic
case.
\begin{description}
\item[PACS numbers]
04.50.Kd
\item[Key Words]
Teleparallel Gravity, Braneworld Gravity
\end{description}
\end{abstract}
\vspace{2cm}

\end{frontmatter}

\section{introduction}

A few years after the introduction of general relativity (GR),
Einstein proposed another gravitational theory called teleparallel
gravity (TG)  in an attempt to unify gravity with electromagnetism
\cite{Ein30}. In this theory he considered a spacetime with zero
curvature but nonzero torsion usually called teleparallel or
Weitzenbock spacetime. This property was in contrast to general
relativity in which the Riemannian spacetime had curvature with
vanishing torsion. Einstein was not able to achieve his primary aim,
as teleparallel gravity was not able to give the unification of
forces. Moreover, the general teleparallel theory also was not
invariant under local Lorentz transformations \cite{Haya79,FT}. By
insisting on restoring the local Lorentz invariance, the theory
becomes equivalent to general relativity, with empirically
indistinguishable results, therefore it is called teleparallel
gravity equivalent to general relativity (TEGR) \cite{Ald10,LI}.
Nowadays, TG and GR are considered to be special cases of the more
general gauge theory of gravity called Poincare gauge theory (PGT)
in which both torsion and curvature are present. In PGT the
spacetime has a Riemann-Cartan structure and both the mass and the
spin of the matter act as the sources of gravitational interaction.
It offers the most realistic and satisfying theory of gravity
\cite{Blag}.

On the other hand the attempt to overcome the problems in GR led to
the born of variety of different proposals, see for example
\cite{NO1,NO2} for reviews. A class of such theories modifies
gravity by considering higher dimensions. The origin of the idea is
a work by Kaluza and Klein \cite{KK} in which only one compact extra
dimension has been considered. Through this assumption, they could
unify gravity and electromagnetism. More recently, works by Randall
and Sundrum led to a simple viable and cosmologically satisfying
braneworld model.  Their second model, RS II, has a single brane
embedded in an infinite bulk. Although in the RS II setup, the extra
dimension is not compact, the 4-D gravity is recovered on the brane
\cite{RS2}. Effectively in this setup the gravity is localized on
the brane by the bulk curvature. The success of this model has
attracted worldwide attention to extra dimensions and many
researches and developments have been so far done in this area
\cite{RSG,RSC}.

Through this paper, we try to revive this five dimensional model in
the context of a 5D TEGR background with the hope of finding
probable intuitive differences between the two gravitational
theories of TEGR and GR. The reason that we choose RS II as the
framework is the existence of the brane as a boundary surface which
separates the two regions of the bulk. This gives an opportunity to
analyze possible difference in the effective 4D gravity as the
junction conditions which connect the 5D quantities to the
stress-energy tensor of the brane, are found to be different in
teleparallel gravity as shown in \cite{BAN}. Another important
question is that whether 4D local Lorentz invariance is recovered on
the brane. This is not a straightforward question in teleparallel
gravity as symmetric and gauge properties of this theory are
fundamentally different from general relativity. 5D models based on
teleparallel gravity are also studied in Refs. \cite{BNO,GLT}.

The structure of the paper is as follows: after introducing
notations and basic definitions in section II, we consider the case
of a Minkowski brane embedded in an Anti de Sitter bulk in section
III. This simple construction allows us to study the linearized
gravity and weak-field limit on the brane effectively. For a more
realistic braneworld setup, we study the case of a
Friedmann-Robertson-Walker brane in a AdS bulk in section IV.

\section{Notation and definitions}

Throughout this paper the capital middle Latin letters $M,N,...$ run
over $0,\,1,\,2,\,3,\,5$ and label spacetime coordinates. Lower case
Latin letters from the beginning of the alphabets $a,b,...$ run over
$0,1,2,3,5$ and label tangent space coordinates. The Greek indices
$\mu,\nu,..$ run over $0,1,2,3$ and refer to the 4D spacetime
coordinates. Finally lower case Latin letters $i,j,...$ run run over
$0,1,2,3$ and refer to the 4D tangent space coordinates.

In teleparallel gravity one considers a set of (pseudo)-orthogonal
D-vectors (D is the number of spacetime's dimensions) which form a
basis in the tangent space on every point of the manifold
$\textbf{e}_{i}\,.\,\textbf{e}_{j}=\eta_{ij}$. This bases are called
tetrads in four dimensions (or pentads in 5D) and relate the
manifold and Minkowski metrics through the relation
\begin{equation}
g_{\mu\nu}=\eta_{ij}e_{\mu}^{i}e_{\nu}^{j}
\end{equation}
The inverse of the tetrad is  defined by the relation
$e_{i}^{\mu}e_{\mu}^{j}=\delta_{i}^{j}$. Here tetrads are the
dynamical variables of the theory. In TEGR, the spin connection
which determines the rule of the parallel transport in the tangent
space, is assumed to be zero. The vanishing spin connection means
that there is no spinor field in the theory. The weitzenb\"{o}ck
connection is defined as \cite{Weitz23}
\begin{equation}
\Gamma^{\rho}_{\,\,\,\,\mu\nu}=e^{\rho}_{i}\partial_{\nu}e^{i}_{\mu}
\end{equation}
which unlike Livi-civita connection is not symmetric on its second
and third indices. Curvature can be defined with respect to spin
connection and torsion with respect to vierbeins and spin
connection. Since the spin connection in this theory is zero,
curvature is also turned out to be zero and then the torsion tensor
is
\begin{equation}
T^{\rho}_{\,\,\,\,\mu\nu}\equiv
e_{i}^{\rho}(\partial_{\mu}e_{\nu}^{i}-\partial_{\nu}e_{\mu}^{i})\,.
\end{equation}
Contorsion tensor which denotes the difference between Livi-civita
and Weitzenb\"{o}ck connections is
\begin{equation}
K^{\mu\nu}_{\quad\rho}=-\frac{1}{2}(T^{\mu\nu}_{\quad\rho}
-T^{\nu\mu}_{\quad\rho}-T_{\rho}^{\,\,\,\,\mu\nu})
\end{equation}
and the superpotential tensor is defined as
\begin{equation}
S^{\,\,\,\,\mu\nu}_{\rho}=\frac{1}{2}(K^{\mu\nu}_{\quad\rho}
+\delta^{\mu}_{\rho}T^{\alpha\nu}_{\quad\alpha}-\delta^{\nu}_{\rho}
T^{\alpha\mu}_{\quad\alpha})\,.
\end{equation}
In correspondence with Ricci scalar, one can define torsion scalar
\begin{equation}
T=S^{\,\,\,\,\mu\nu}_{\rho}T^{\rho}_{\,\,\,\,\mu\nu}
\end{equation}
The gravitational action in TEGR is
\begin{equation}
I=\frac{1}{16\pi G}\int dx\,|e|\, T
\end{equation}
where $|e|$ is the determinant of $e^{a}_{\mu}$ and from the
relation (1), one can easily finds that it is equal to $\sqrt{-g}$.
Variation of the above action with respect to vierbeins gives the
field equations in TEGR
\begin{equation}
e^{-1}\partial_{\mu}(ee_{i}^{\rho}S^{\,\,\,\,\mu\nu}_{\rho})
-e_{i}^{\lambda}T_{\,\,\,\,\mu\lambda}^{\rho}S^{\,\,\,\,\nu\mu}_{\rho}
+\frac{1}{4}e_{i}^{\nu}T=4\pi G e_{i}^{\rho}\Xi^{\,\,\nu}_{\rho}
\end{equation}
where $\Xi^{\,\,\nu}_{\rho}$ is the energy-momentum tensor.\\
 The Lagrangian of TEGR (Eq. (7))  is consisted of just the
torsion scalar $T$. Torsion scalar differs only by a total
divergence term from Ricci scalar $R$ of general relativity,
$R=-T-2\nabla_{\mu}T_{\nu}^{\,\,\mu\nu}$. $R$ is local Lorentz
invariant but the total divergence term
$\nabla_{\mu}T_{\nu}^{\,\,\mu\nu}$ is not, which means that contrary
to $R$, the torsion scalar cannot be local Lorentz invariant . The
$\nabla$ operator here is with respect to the Levi-civita
connection. This total divergence term will vanish in an action
integral and the field equations will become equal to that of GR. In
five dimensional TEGR with RS II as the framework, to see if the
effective theory also remains invariant under local Lorentz
transformations, one should study the behavior of transformed
quantities when one projects them on the brane. Under such
transformations in the tangent space, vierbein and torsion tensor
transform as
\begin{equation}
e^{a}_{\,M}\mapsto\Lambda^{a}_{\,b}e^{b}_{\,M}
\end{equation}

\begin{equation}
T^{M}_{\,NQ}\mapsto
T^{M}_{\,NQ}+\Lambda_{a}^{\,b}e_{b}^{\,M}(e^{c}_{\,Q}\partial_{N}\Lambda^{a}_{\,c}-e^{c}_{\,N}\partial_{Q}\Lambda^{a}_{\,c})
\end{equation}
therefore the total divergence term in the five dimensional
Lagrangian becomes
\begin{equation}
\nabla_{N}T_{M}^{\,\,NM}\mapsto\nabla_{N}T_{M}^{\,\,NM}+\nabla_{N}(\Lambda_{a}^{\,c}\partial^{N}\Lambda^{a}_{\,c}-
\Lambda_{a}^{\,b}e_{b}^{\,N}e^{c}_{\,M}\partial^{M}\Lambda^{a}_{\,c})
\end{equation}
It is clear that the term which involves the Lorentz transformation
tensor is completely separated from the untransformed torsion tensor
and is still a total divergence. To obtain the effective Lagrangian
on the brane, one should integrate the 5-D Lagrangian with respect
to the extra dimension $y$. If the integration of the second term
with respect to $y$ vanishes, the effective Lagrangian on the brane
remains invariant. This depends on the specific geometry of the bulk
and the brane and we will examine this for the cases of Minkowski
and FRW branes in the following sections. However, since the
transformed part is still a total divergence term, when integrating
over whole spacetime, it will definitely vanishes and leaves the 4-D
effective field equations invariant independent of goometry. To sum
up, although in general, the 4-D effective Lagrangian in TEGR is not
equivalent with that of GR due to
different junction conditions, but the theory remains local Lorentz invariant on the level of field equations.\\

Here we briefly present the main results of Ref \cite{BAN}. In that
paper, starting from a 5D Randall-Sundrum setup, the effective 4D
field equations on the brane were derived by projecting all the 5D
geometrical quantities using the equivalent of Gauss-Codacci
equations in teleparallel gravity. The energy-momentum tensor can be
considered as
\begin{equation}
\Xi_{MN}=-\Lambda_{5}g_{MN}+\delta(y)\Omega_{MN}
\end{equation}
where $\Lambda_{5}$ is the five dimensional cosmological constant,
$\lambda$ is the brane tension and $\Omega_{MN}$ is the matter
stress-energy tensor of the brane. To find the induced field
equations on the brane, using the procedure first introduced in
\cite{SMS}, we project all the five dimensional quantities by using
the projection pentad \cite{BAN}
\begin{equation}
h^{M}_{a}=e^{M}_{a}-n^{M}n_{a}\,.
\end{equation}
this projection tensor when acts on a vector, will project it on the
brane and turns the tangent indices into coordinate and vice versa.

In RS II, brane is actually a border which divides bulk into two
regions. Going from one side of the brane to the other, will cause
some discontinuities in our physical quantities. To encounter these
discontinuities one needs junction conditions. Expressing the pentad
field as a distribution and requiring that the connection also be a
distribution,  we reach the first junction condition
\begin{equation}
[\,e_{M}^{a}\,]=0
\end{equation}
This means that the pentad is continuous across the brane.
Expressing other geometrical quantities in the same way and using
the five dimensional field equations, we reach the second junction
condition which guaranties the geometry of the theory remains
well-defined. This relates the jump of the five dimensional
superpotential tensor across the brane,  to the matter content on
the brane as \cite{BAN}
\begin{equation}
e_{a}^{O}\,[S^{~MN}_{O}]\,n_{M}=4\pi G\,\,\Omega_{a}^{N}
\end{equation}
Substituting projected quantities on the brane and using the above
junction conditions and imposing the $Z_{2}$-symmetry, we obtain the
induced field equations on the brane \cite{BAN}
\begin{equation}
^{(4)}F_{a}^{N}=-\Lambda_{5}h_{a}^{N}+(4\pi
G_{5})^{2}\Pi_{a}^{N}+E_{a}^{N}
\end{equation}
where we have defined
\begin{eqnarray}
\nonumber\Pi_{a}^{N}=&-&\frac{3}{4}h_{O}^{b}\Omega_{b}^{N}\Omega^{O}_{a}+\frac{3}{8}h_{a}^{O}\Omega\Omega_{O}^{N}
+\frac{1}{32}h_{a}^{N}\Omega_{b}^{O}\Omega^{b}_{O}+\frac{1}{32}h_{a}^{N}\Omega^{2}\\
&+&\frac{1}{4}\Phi^{2}(1+L_{M}J^{M})\delta_{O}^{N}\Omega\Omega_{a}^{O}
+\frac{1}{4}\Phi^{2}(1+L_{M}J^{M})\delta_{a}^{N}\Omega^{2}\nonumber\\
\end{eqnarray}
and
\begin{eqnarray}
\nonumber
E_{a}^{N}&=&n^{O}n_{a}\partial_{M}(S^{\,\,\,MN}_{O})+S^{\,\,\,MN}_{O}(n^{O}\partial_{M}n_{a})
+S^{\,\,\,MN}_{O}(n_{a}\partial_{M}n^{O})+h_{a}^{O}S^{\,\,\,MN}_{O}(n^{b}\partial_{M}n_{b})\\
\nonumber&~&+\Bigg[n^{M}n_{O}n_{b}n^{c}e_{d}^{N}+n^{N}n_{O}n_{d}n^{c}e_{b}^{M}+n^{M}n_{b}n_{d}n^{N}e_{O}^{c}\\
&~&-n^{c}n_{O}e_{b}^{M}e_{d}^{N}-n^{M}n_{b}e_{O}^{c}e_{d}^{N}
-n^{M}n_{O}n_{b}n^{c}n^{N}n_{d}\Bigg]
S^{\,\,\,bd}_{c}\partial_{M}(hh_{a}^{O})
\end{eqnarray}

where pentad and the inverse pentad are given by
\[e_{\mu}^{i}(x,y)=\Bigg(\matrix{ e_{\alpha}^{a} & 0\cr e_{\alpha}^{.5} &
e_{5}^{.5}}\Bigg)\]

\[e_{i}^{\mu}(x,y)=\Bigg(\matrix{ e^{\alpha}_{a} & e^{\alpha}_{.5}\cr 0 &
e^{5}_{.5}}\Bigg)\]

respectively and we have defined
\begin{equation}
e_{\mu}^{.5}=L_{\mu}\Phi \quad e_{5}^{.5}=\Phi\quad ,\quad
e_{i}^{5}=-h_{i}^{\mu}L_{\mu} \quad e_{.5}^{5}=\Phi^{-1}
\end{equation}
and $J_{i}=\Phi^{-1}\partial_{i}\Phi$. A `$.$' in front of an index
refers that it is a tangent space index.

\section{Setup of the Randall-Sundrum model in TEGR}
\subsection{Five dimensional geometry setup in TEGR}
 In this section we wish to investigate a RS type
scenario in the context of teleparallel gravity. Here the
gravitational interactions are described by torsion instead of
curvature. One of the fundamental assumptions of the original RS
model was that the metric ansatz obeys the 4-D Poincare invariance.

Anti de Sitter space is the maximally symmetric solution of
Einstein's equations with an attractive cosmological constant. It
has constant negative scalar curvature. The $AdS_{5}$ metric usually
takes the form
\begin{eqnarray}
\nonumber ds^{2}&=&dy^{2}+e^{2A(y)}\eta_{\mu\nu}dx^{\mu}dx^{\nu}\\
&=&dy^{2}+e^{2A(y)}\Big[dx^{2}+dy^{2}+dz^{2}-dt^{2}\Big]
\end{eqnarray}
This will be a $AdS_{5}$ metric if we set $A=\pm b$. $K=-b^{2}$ is
the constant negative curvature of $AdS_{5}$ space in general
relativity. By a simple coordinate transformation in the form of
$\upsilon = \frac{e^{\mp by}}{b}$ the above metric can be
transformed into:
\begin{equation}
ds^{2}=\frac{1}{b^{2}\upsilon^{2}}\Big[d\upsilon^{2}+dx^{2}+dy^{2}+dz^{2}-dt^{2}\Big]
\end{equation}
we can see that the $AdS_{5}$ metric is indeed conformally flat as
\begin{equation}g_{\mu\nu}=\Omega^{2}\eta_{\mu\nu}
\end{equation}
 where
\begin{equation}
\Omega^{2}=e^{\pm 2by}
\end{equation}
This coordinate system is usually called the 'stereographic
coordinates' in the literature. In Teleparallel gravity we have
\cite{TG}
\begin{equation}
g_{\mu\nu}=e_{\,\,\,\mu}^{i}e_{\,\,\,\nu}^{j}\eta_{ij}
\end{equation}
where $e_{\mu}^{i}$ is the tetrad field. By using equation $(22)$
and $(24)$ we see that the tetrad field for the $AdS_{5}$ space in
the stereographic coordinate is \cite{TG}
\begin{equation}
e_{\,\,\,\mu}^{i}=\Omega\,\delta_{\mu}^{i}
\end{equation}
Weitzenbock connection has the form
\begin{equation}
\Gamma_{\,\,\,\mu\nu}^{\rho}=e^{\,\,\,\rho}_{i}\partial_{\mu}e_{\,\,\,\nu}^{i}
\end{equation}
Using the $AdS_{5}$ tetrad $(6)$ we have
\begin{equation}
\Gamma_{\,\,\,\mu\nu}^{\rho}=\delta_{\nu}^{\rho}\partial_{\mu}\ln\Omega
\end{equation}
and the torsion will be
\begin{equation}
T_{\,\,\,\mu\nu}^{\rho}=\delta_{\nu}^{\rho}\partial_{\mu}\ln\Omega-\delta_{\mu}^{\rho}\partial_{\nu}\ln\Omega
\end{equation}
So in this particular coordinate system the torsion has this simple
form. Substituting for $\Omega$ from (23) shows us that the
$AdS_{5}$ space in theleparallel gravity has constant negative
torsion scalar and its curvature is identically zero. We can also
easily show that the $AdS_{5}$ space is indeed the solution of the
teleparallel field equation (8) with a negative cosmological
constant where
\begin{equation}
\Xi^{\,\,N}_{a}=-\Lambda_{5}e_{a}^{N}
\end{equation}

By introducing the brane into this setup some restrictions will be
imposed in the form of the warp factor. In order to have a well
defined and acceptable Cauchy development, the warp factor in the
presence of the brane should be
\begin{equation}
e^{2A(y)}=e^{-2b|y|}
\end{equation}
This space is usually called the $AdS_{5}/Z_{2}$ space and is a
choice which is bounded everywhere and  unlike the unbounded
$e^{2A(y)}=e^{2b|y|}$ it has a well developed Cauchy problem.

Using this warp factor and turning the attention to the original
$AdS_{5}$ line element (20), one can see that the simplest pentad
field in this setup is given by
\begin{equation}
e_{\,\,\,M}^{a}=\texttt{diag}\Big(e^{-b|y|},e^{-b|y|},e^{-b|y|},e^{-b|y|},1\Big)
\end{equation}
We now proceed to calculate the torsion and superpotential with this
tetrad. We note that
\begin{equation}
\frac{d|y|}{dy}=\Theta(y)-\Theta(-y)=\epsilon(y)\,\,\,\,\, ,
\,\,\,\,\,\, \frac{d^{2}|y|}{dy^{2}}=2\delta(y)
\end{equation}
where $\theta(y)$ is the Heaviside distribution which is defined as
follows: it is equal to $+1$ if $y>0$, $0$ if $y<0$ and
indeterminate if $y=0$. It has the following properties
\begin{equation}
\Theta^{2}(y)=\Theta(y) \,\,\,\,\ ,
\,\,\,\,\Theta(y)\Theta(-y)=0\,\,\,\,,\,\,\,\,\frac{d}{dy}\Theta(y)=\delta(y)
\end{equation}
where $\delta(y)$ is the Dirac distribution.\\
The non-zero components of the torsion are
\begin{equation}
T_{\,\,\,50}^{0}=-b(\Theta(y)-\Theta(-y))\,\,\, ,
\,\,\,T_{\,\,\,51}^{1}=T_{\,\,\,52}^{2}=T_{\,\,\,53}^{3}=-b(\Theta(y)-\Theta(-y))
\end{equation}
and the non-zero components of the superpotential are
\begin{equation}
S_{0}^{\,\,\,50}=S_{1}^{\,\,\,51}=S_{2}^{\,\,\,52}=S_{3}^{\,\,\,53}=\frac{3}{2}b(\Theta(y)-\Theta(-y))
\end{equation}
So the torsion scalar will be
\begin{eqnarray}
T&=&-6b^{2}\Big(\Theta(y)-\Theta(-y)\Big)^{2}=-6b^{2}\Big(\Theta^{2}(y)+\Theta^{2}(-y)-2\Theta(y)\Theta(-y)\Big)\nonumber\\
&=& -6b^{2}\Big(\Theta(y)-\Theta(-y)\Big)=-6b^{2}
\end{eqnarray}
So the $AdS_{5}$ space in teleparallel gravity indeed has constant
negative scalar torsion.

If we denote the left hand side of the teleparallel field equation
(8) by $F_{a}^{N}$, then we have
$$F^{0}_{.0}=e^{-3b|y|}\,6b^{2}-e^{-b|y|}\,6b\,\delta(y)$$
\begin{equation}
F^{1}_{.1}=F^{2}_{.2}=F^{3}_{.3}=e^{-b|y|}\,6b\,\delta(y)-e^{-3b|y|}\,6b^{2}
\, ,\,F^{5}_{.5}=-e^{-b|y|}\,6b^{2}
\end{equation}
where a `.' denotes the tangent space indices. From these equations
we see that as well as having a cosmological constant in the bulk,
there should be an additional energy momentum tensor on the brane
(with a delta function). The complete energy - momentum tensor which
supports this particular form of tetrad with this specific warp
factor then should be
\begin{equation}
\Xi^{\,\,N}_{a}=-\Lambda_{5}e_{a}^{N}+\lambda e_{a}^{N} \delta(y)
\end{equation}
where $\lambda$ is the cosmological constant induced on the brane or
the brane tension. Using the teleparallel field equations we get
\begin{equation}
\lambda=\frac{6b}{\kappa^{2}_{5}}\,\,\,\,\, , \,\,\,\,
\Lambda_{5}=\frac{-6b^{2}}{\kappa^{2}_{5}}
\end{equation}
So we also have
\begin{equation}
\kappa^{2}_{5}\lambda^{2}+6\Lambda_{5}=0
\end{equation}
The presence of this additional energy-momentum tensor entails the
presence of a new matter field $\lambda$ which is localized to the
$y = 0$ region which is associated with the brane localized there.
In summary in this section we have constructed a 5-D AdS geometry in
teleparallel gravity. Introducing the brane in this setup will
induce some restrictions on the coefficient of the AdS pentad.

\subsection{Effective 4-D action}
The 5-D gravitational action in the RS setup is
\begin{equation}
S_{grav}=-\frac{1}{\kappa^{2}_{5}}\int d^{4}x\, \int^{\pi}_{0} d\phi
r_{c}|e| (-\Lambda_{5}+T)
\end{equation}
where we have introduced $y=r_{c}\phi$ and $\phi$ goes from $0$ to
$2\pi$. $r_{c}$ essentially is the "compactification radius" of the
extra dimensional circle. In order to study the effective 4-D
action, we begin by considering small fluctuations around the 4-D
tetrad. This can be achieved by replacing the Minkowski metric in
(20) by a four dimensional metric $\bar{g}_{\mu\nu}(x)$ where
\begin{equation}
\bar{g}_{\mu\nu}(x)=\eta_{\mu\nu} + \bar{h}_{\mu\nu}
\end{equation}
so we have
\begin{equation}
ds^{2} = dy^{2} + e^{-2b|y|} (\eta_{\mu\nu} +
\bar{h}_{\mu\nu})dx^{\mu}dx^{\nu}
\end{equation}
In terms of the tetrad field, the 4-D fluctuations can be written as
\begin{equation}
\bar e_{\,\,\,\mu}^{i}(x)=\delta_{\mu}^{i}+\bar h_{\,\,\,\mu}^{i}(x)
\end{equation}
where $\bar{e}_{\,\,\,\mu}^{i}$ is the 4-D tetrad.

 up to the first order in perturbations, the non-zero  torsion components of the tetrad
$(43)$ are

$$T_{\,\,\,50}^{0}=-b(1+h^{.0}_{0}+h_{.0}^{0})\,\,\, , \,\,\,$$
$$T_{\,\,\,51}^{1}=-b(1+h^{.1}_{1}+h_{.1}^{1})\,\,\,,\,\,\,T_{\,\,\,52}^{2}=-b(1+h^{.2}_{2}+h_{.2}^{2})$$
$$T_{\,\,\,53}^{3}=-b(1+h^{.3}_{3}+h_{.3}^{3})$$
\begin{equation}
T_{\,\,\,5\nu}^{\rho}=-b(\delta_{i}^{\rho}h_{\nu}^{i}
+\delta_{\nu}^{i}h_{i}^{\rho})\,\,\,,\,\,\,T_{\,\,\,\mu\nu}^{\rho}
=e^{-b|y|}e^{b|y|}\bar{T}_{\,\,\,\mu\nu}^{\rho}=\bar{T}_{\,\,\,\mu\nu}^{\rho}
\end{equation}
where $\bar{T}_{\,\,\,\mu\nu}^{\rho}$ is the torsion constructed by
the 4 dimensional tetrad $(44)$ . Similarly for the superpotential
we have
$$S^{\,\,\,50}_{0}=\frac{3}{2}b-b(1+h^{.0}_{0}+h_{.0}^{0})\,\,\, , \,\,\,
S^{\,\,\,51}_{1}=\frac{3}{2}b(1+h^{.1}_{1}+h_{.1}^{1})$$
\begin{equation}
\,\,\,,\,\,\,S^{\,\,\,52}_{2}=\frac{3}{2}b(1+h^{.2}_{2}+h_{.2}^{2})\,\,\,,\,\,\,
S^{\,\,\,53}_{3}=\frac{3}{2}b(1+h^{.3}_{3}+h_{.3}^{3}) ,
S^{\,\,\,\mu\nu}_{\rho}= e^{b|y|}\bar{S}^{\,\,\,\mu\nu}_{\rho}
\end{equation}
And finally for the torsion scalar $T$ we have
\begin{equation}
T=-6b^{2}\Big[1+2\texttt{Tr}(h_{\mu}^{i})+2\texttt{Tr}(h^{\mu}_{i})+\sum^{3}_{i=0}(h_{\mu}^{i}\delta_{i}^{\mu})^{2}\Big]+e^{b|y|}\,\bar{T}
\end{equation}
which again $\bar{T}$ is the torsion scalar constructed by (44) and
is purely 4 dimensional and y-independent. If we work in the
Transverse-Traceless gauge then
$\texttt{Tr}(h_{\mu}^{i})=\texttt{Tr}(h^{\mu}_{i})=0$ and up to
first order we have
\begin{equation}
T=-6b^{2}+e^{b|y|}\,\bar{T}\,.
\end{equation}
Note that the fourth term in the RHS of (47) is a term second order
in fluctuation and can be neglected. For the determinant of pentad
we can easily see that there exists the following relation between
5-D determinant constructed by (43) and 4-D determinant constructed
by (44)
\begin{equation}
e=e^{-4b|y|}\bar{e}
\end{equation}
where $\bar{e}$ is $y$-independent. Substituting (48) and (49) in
the action (41), we have
\begin{equation}
S_{grav}=-\frac{1}{\kappa^{2}_{5}}\int d^{4}x\, \int^{\pi}_{0} d\phi
r_{c}e^{-4br_{c}|\phi|}\bar{e}(-6b^{2}+e^{br_{c}|\phi|}\,\bar{T})\,.
\end{equation}
The first term in the above integral is a constant and can be
integrated and absorbed into the cosmological constant term in (41).
The second term gives us the effective 4-D action and effective 4-D
Planck scale. In this case when we consider only the first order
fluctuations, the effective 4-D action is indeed only $\bar{T}$ and
as a result this geometrical setup is equivalent to GR in both
levels of the action (up to the first order) and the field
equations.
\section{FRW brane embedded in AdS bulk}
For the embedding of a (not necessarily static) maximally
3-symmetric geometry in a 5-dimensional bulk, the most general line
element which respects the maximal 3- symmetry is given by
\cite{Manh}

 $$ds^2=-n^2(y,t)dt^2+2c(y,t)dydt$$
\be
 +a^2(y,t)\Bigg[\frac{dr^2}{1-kr^2}+r^2d\Omega^2\Bigg]+b^2(y,t)dy^2
\ee

The most general pentad which gives this geometry and also respect
the fundamental structure of spacetime as given by eq (19) is
\begin{equation}
e^{a}_{M}= \left(
\begin{array}{ccccc}
\sqrt{(\frac{c^2}{b^2}+n^2)} & 0 & 0 & 0 & 0 \nonumber\\
0 & a & 0 &0 &0
 \nonumber\\
0& 0 & a&0&0 \nonumber\\
0& 0 & 0&a&0\nonumber\\
\frac{c}{b}&0&0&0&b\nonumber
\end{array}
\right) \, ,
\end{equation}
However starting from TEGR in 5 dimensions and noting that the 5D
field equations is invariant under local Lorentz transformation of
the pentad, \be e^{a}_{M}\rightarrow \Lambda_{b}^{a}\,\,e^{b}_{M}
 \ee, one can write
most general FRW pentad as without loss of generality
\begin{equation}
e^{A}_{\mu}= \left(
\begin{array}{ccccc}
n & 0 & 0 & 0 & 0 \nonumber\\
0 & a & 0 &0 &0
 \nonumber\\
0& 0 & a&0&0 \nonumber\\
0& 0 & 0&a&0\nonumber\\
0&0&0&0&b\nonumber
\end{array}
\right) \, ,
\end{equation}
 The TEGR Lagrangian is also invariant under general coordinate
transformation. Using this extra freedom, we choose the gauging
$b=1$ which corresponds to the Gaussian normal gauge in general
relativity and effectively fixes the position of the brane in the 5D
geometry. For simplicity we choose $y=0$ as the position of the
brane. Note that setting $b=1$ brings down the number of independent
coefficients of the pentad (54) to two. This is acceptable as the
number of independent metric coefficients in the corresponding
general relativistic setup is also two and we expect our 5D TEGR
theory to posses the same number of degrees of freedom as general
relativity in five dimensions. The transformation which transforms
the pentad (52) to (54) is given by
\begin{equation}
M^{A}_{~B}= \left(
\begin{array}{ccccc}
\frac{n}{\sqrt{(\frac{c^2}{b^2}+n^2)}} & 0 & 0 & 0 & 0 \nonumber\\
0 & 1 & 0 &0 &0
 \nonumber\\
0& 0 & 1&0&0 \nonumber\\
0& 0 & 0&1&0\nonumber\\
\frac{-c}{\sqrt{(\frac{c^2}{b^2}+n^2)}}&0&0&0&1\nonumber
\end{array}
\right) \, ,
\end{equation}
in the $t-y$ plane we have
\begin{equation}
M^{A}_{~B}= \left(
\begin{array}{ccccc}
\frac{n}{\sqrt{(\frac{c^2}{b^2}+n^2)}} & 0  \nonumber\\
\frac{-c}{\sqrt{(\frac{c^2}{b^2}+n^2)}}&1\nonumber
\end{array}
\right) \, ,
\end{equation}
defining $\cos(\theta)=\frac{\sqrt{(\frac{c^2}{b^2}+n^2)}}{n}$, we
see that
\begin{equation}
M^{A}_{~B}= \left(
\begin{array}{ccccc}
\frac{1}{\cos(\theta)} & 0  \nonumber\\
-\tan(\theta)&1\nonumber
\end{array}
\right) \, ,
\end{equation}
which is exactly the transformation between a non-orthogonal basis
and an orthogonal one. Note that this is not a local Lorentz
transformation as it does not satisfy the condition $\Lambda^{T}\eta
\Lambda=\eta$ where $\eta$ is the Minkowski metric of the tangent
space.

The fact that these two pentads which are not connected to each
other through a local Lorentz transformation, both describe the same
geometry with the same dynamics, implies that there are extra hidden
symmetries in the this setup of the theory that can be used to
reduce the number of independent degrees of freedom to that of
general relativity in 5 dimensions.

The situation of the effective 4D dynamics on the brane is quite a
different matter. As the pentad coefficients in (54) are not
separable functions of $t$ and $y$, finding of the 4D effective
dynamics is not a straightforward question like the Minkowski case.
The 4D brane dynamics is derived from the bulk quantities through
the junction conditions. As the junction conditions in the
teleparallel braneworld gravity differ from general relativity, one
could expect some modifications in 4 dimensions. The induced field
equation derived in \cite{BAN} shows this feature. Using the FRW
pentad (54) with $b=1$, we derive the torsion, contortion and the
superpotential tensors of the 5D background. The torsion scalar then
reads
\begin{equation}
T=\frac{-6a'n'}{an}+\frac{6\dot{a}^2}{a^2n^2}-\frac{6a'\,^2}{a^2}
\end{equation}
substituting in the 5D teleparallel field equations we get
\begin{eqnarray}
F_{.5}^{0}=F_{.0}^{5}=-\frac{3}{2}\frac{\dot{a}'}{a^4n^3}+\frac{3}{2}\frac{n'\dot{a}}{a^4n^4}=0
\end{eqnarray}
\begin{eqnarray}
F_{.0}^{0}=-3\frac{a'^2}{a^2}-3\frac{a''}{a}+\frac{3\dot{a}^2}{a^2n^2}=4\pi
G \Big(\rho(t)\delta(y)+\Lambda_{5}\Big)
\end{eqnarray}
\begin{eqnarray}
F_{.1}^{1}
=-2a''a+a'^2+\frac{2aa'n'}{n}+\frac{n''a^2}{n}-\frac{\dot{a}^2}{n^2}+\frac{3a\dot{a}\dot{n}}{n^3}
-\frac{2a\ddot{a}}{n^2}=4\pi G \Big(p(t)\delta(y)+\Lambda_{5}\Big)
\end{eqnarray}
\begin{eqnarray}
F_{.5}^{5}
=3\frac{a'^2}{a^2}+3\frac{a'n'}{an}-\frac{3\dot{a}^2}{a^2n^2}+\frac{3\dot{a}\dot{n}}{n^3}+3\frac{\ddot{a}}{a}=4\pi
G \Lambda_{5}
\end{eqnarray}
Here `dot' denotes derivative with respect to time $t$ and a `prime'
denotes derivative with respect to $y$. Note that these are exactly
the same 5D equations as in general relativity \cite{BMW,RVSD}. This
is expected as we are working in the teleparallel equivalent of
general relativity in 5 dimensions. Any possible difference in
effective 4D theory is then coming from the different junction
conditions. To solve these equations, we first solve them in the
bulk and then impose the junction conditions at the brane. Equation
(59) can be simplified to
\begin{equation}
\frac{\dot{a}'}{a}=\frac{n'}{n}
\end{equation}
The non-trivial solution to this equation is
\begin{equation}
\dot{a}=D(t) n
\end{equation}
where $D(t)$ is a function that depends only on time and is
independent of $y$. Substituting this solution in equation (60) and
equating it to $-6b^2$ in the bulk gives
\begin{equation}
\frac{D(t)}{a^2}-\frac{a'^2}{a^2}-\frac{a''}{a}=-2b^2
\end{equation}
the solution is

 \bea a \left( y,t \right) = \frac{\sqrt{2}}{2}\,{\frac
{\sqrt {{{\rm e}^{-2\,b|y|}} \left( -{{\rm
e}^{-2\,b|y|}}D(t)+b\,{{\rm e}^{-4\,b|y|}}{\it A(t)}-b\,{ \it
B(t)}\right) }}{b\,{{\rm e}^{-2\,b|y|}}}}\eea

where $A(t)$ and $B(t)$ are arbitrary t-dependant integration
functions.

Imposing the junction conditions (15) on the brane we get
\begin{equation}
\frac{3}{2}\frac{a'(0,t)}{a(0,t)}=4\pi G \rho
\end{equation}
\begin{equation}
\frac{a'(0,t)}{a(0,t)}+\frac{n'(0,t)}{n(0,t)}= 4\pi G p
\end{equation}
Substituting (68) and (69) in (71) and (72) yields
\begin{equation}
\frac {-3{b}^{2} \Big( A(t)+B(t) \Big) }{  2\, A(t)b-2\,B(t)
b-2\,D(t)}=4\pi G \rho(t)
\end{equation}

$${\frac { \left(  \left( B(t) +A(t)   \right) \dot{D}
\left( t \right) + \left( - B(t)  b-2\,D \left( t \right) +3\,A(t) b
\right) \dot{B(t)}   \right) {b}^{2}}{ \Big( -D \left( t \right)
+A(t)  b-B(t)  b
 \Big)  \left( -\dot{D} \left( t \right)
  +\dot{A(t) }b-  \dot{B(t)}b \right) }}$$
$$
+\frac{ b^2\Bigg(
  \dot{A(t)}\left( -3\,
B(t)  b-2\,D \left( t \right) +A(t)
  b \right)\Bigg)}{\left( -D
\left( t \right) +A(t)  b-B(t)  b
 \right)  \left( -\dot{D} \left( t \right)
  +\dot{A(t) }   b-  \dot{B(t)}  b \right) }
  $$
\begin{equation}
 +\frac{-3{b}^{2} \Big( A(t)+B(t) \Big) }{  2\, A(t)b-2\,B(t)
b-2\,D(t)}=4\pi G p(t)
\end{equation}

By assigning appropriate $\rho(t)$ and $p(t)$, these two equations
along with the 5-5 equation (62) when evaluated at the brane, will
fully specify all tetrad coefficients and with that, the dynamics of
a FRW brane embedded in an AdS bulk will be fully determined.

Substituting (58) in the action (41) in the case of a FRW brane, and
explicitly evaluating the integral over $y$, will give us effective
4D action as
\begin{eqnarray}
\L_{eff}&=&\Phi\arctan\Bigg[\frac{\dot{D}-2\dot{A}b}{\sqrt{-4b^2\dot{B}\dot{a}-\dot{D}^2}}\Bigg]
+\Psi\arctan\Bigg[\frac{D-2Ab}{\sqrt{4b^2BA-D^2}}\Bigg]\nonumber\\
&\quad&+\Upsilon\ln\Bigg[\frac{D-bA-bB}{\dot{D}-\dot{A}b+\dot{B}b}\Bigg]
\end{eqnarray}
where
\begin{equation}
\Phi\equiv\frac{24b\Big(B\dot{A}D+A\dot{B}D-2A\dot{D}B\Big)\Big(b^2\dot{B}\dot{A}
-\frac{1}{4}D^2\Big)}{b^2\dot{A}^3B^2-2b^2AB\dot{B}\dot{A}-D^2\dot{B}\dot{A}+D\dot{D}BA}
\end{equation}
\begin{eqnarray}
\Psi&\equiv&
24b\Bigg\{\frac{-1}{2}b^2A^2D^2B^2+\frac{1}{2}D^4A\dot{B}-\frac{1}{2}D^3A\dot{B}\nonumber\\
&\quad&+b^2BD^2A^2\dot{B}-2b^2A^3DB\dot{B}+b^2A^2B^2\dot{A}\dot{D}\nonumber\\
&\quad&-\frac{1}{2}ABD^3\dot{A}\dot{D}+\frac{1}{4}ABD^2\dot{A}\dot{D}-\frac{1}{2}b^2A^2D^2\dot{B}^3\dot{D}
\nonumber\\
&\quad&+b^2A^2B\dot{B}\dot{D}-\frac{1}{2}AD^3\dot{B}\dot{D}+\frac{1}{4}AD^2\dot{B}\dot{D}+\frac{1}{2}ABD^2\dot{D^2}\Bigg\}\nonumber\\
&\quad&
\times\Bigg\{b^2\dot{A}^3B^2-2b^2AB\dot{B}\dot{A}-D^2\dot{B}\dot{A}+D\dot{D}BA\Bigg\}^{-1}
\end{eqnarray}
\begin{equation}
\Upsilon\equiv\frac{24b\Bigg[\frac{1}{4}(b^2A\dot{B}-\frac{1}{2}D\dot{D}+b^2AB)(A\dot{B}-B\dot{A})\Bigg]}
{b^2\dot{A}^3B^2-2b^2AB\dot{B}\dot{A}-D^2\dot{B}\dot{A}+D\dot{D}BA}
\end{equation}
As we can see, the effective Lagrangian in the FRW case is not the
same as in GR and as a result the 4-D cosmological dynamics will be
different from GR. For a quick glance at the practical results of
the model and study quantitative differences, we consider a simple
inflationary universe when the exponential expansion is driven by a
scalar field. Using the procedure above and eq (16), the Friedmann
equation on the brane will be \cite{BNO}
\begin{equation}
\frac{\dot{a}^2}{a^2}=\frac{1}{3}\Bigg[-\Lambda_{4}+8\pi G
\rho+\frac{1}{4}(4\pi G)^2(11-60\omega+93\omega^2)\rho^2\Bigg]\\
\end{equation}\\
where $\omega=\frac{p}{\rho}$ is the equation of state parameter of
the matter confined to the brane. The scalar field, $\phi$ which
drives the inflation has energy density and pressure
\begin{eqnarray}
\rho=\frac{1}{2}\dot{\phi}^{2}+V
\,\,\,,\,\,\,p=\frac{1}{2}\dot{\phi}^{2}-V
\end{eqnarray}
respectively where $V(\phi)$ is the inflation potential. We define
the slow-roll parameters as usual
\begin{equation}
\epsilon\equiv\frac{{M_{4}}^{2}}{4\pi}\bigg(\frac{H'}{H}\bigg)^{2}
\,\,\,,\,\,\,\eta\equiv\frac{{M_{4}}^{2}}{4\pi}\bigg(\frac{H''}{H}\bigg)
\end{equation}
In the slow-roll regime we have
\begin{eqnarray}
\frac{1}{2}\dot{\phi}^2\ll V(\phi)\,\,\,,\,\,\, 3H\dot{\phi} \simeq
-V'(\phi)
\end{eqnarray}
We assume a chaotic type potential for the inflaton field
\begin{equation}
V(\phi)=\frac{1}{2}m^2\phi^2
\end{equation}
Substituting we have
\begin{equation}
\epsilon=\pi G \Big(\frac{V'}{V}\Big)^2\Big(\frac{1+1312\pi G
V}{1+656\pi G V}\Big)^2
\end{equation}
\begin{equation}
\eta=\pi G \Big(\frac{V''}{V}\Big)\small\Bigg[\frac{26896 (4\pi
G)^2V^2+246 (4\pi G)V+8\pi G-\frac{V'^2}{V''V}}{(1+(4\pi G) 164
V)^2}\Bigg]
\end{equation}
where here a prime denotes the differentiation with respect to the
argument. Comparing these relations with the corresponding results
in general relativity in ref \cite{Mar}, we find that contrary to GR
case, here the slow-roll parameters are enhanced by brane
modifications.

Other inflation parameters can then be derived using standard
procedures. The scalar spectral index will be

$$n_{s}=1-6\pi G \Big(\frac{V'}{V}\Big)^2\Big(\frac{1+1312\pi G
V}{1+656\pi G V}\Big)^2$$
\begin{equation}
~~~~~~~~~~~~~~+2\pi G \Big(\frac{V''}{V}\Big)\small\Bigg[\frac{26896
(4\pi G)^2V^2+246 (4\pi G)V+8\pi G-\frac{V'^2}{V''V}}{(1+(4\pi G)
164 V)^2}\Bigg]
\end{equation}
The amplitude of scalar and tensor perturbations then are
\begin{equation}
A^{2}_{S}=\frac {9}{25}\,\frac{{\Big[\frac{8}{3}\,\pi \,GV+{\frac
{656}{3}}\,{\pi }^{2}{G}^{2}{V }^{2}\Big]}^{6}}{ V'^{2} }
\end{equation}

and

\begin{equation}
A^{2}_{T}={\frac {1}{1600}}\,\frac{{\Big[\frac{8}{3}\,\pi
\,GV+{\frac {656}{3}}\,{\pi }^{2}{G}^{2} {V}^{2}\Big]}^{2}}{{\pi
}^{5}{G}^{4}}
\end{equation}
Respectively, then the tensor-to-scalar ratio will be
\begin{equation}
\frac{A^{2}_{T}}{A^{2}_{S}}={\frac
{1}{576}}\,\frac{V'^{2}}{{\Big[\frac{8}{3}\,\pi \,GV+{\frac
{656}{3}}\,{\pi }^{2}{G}^{2}{V}^{2}\Big]}^{4}{ \pi }^{5}{G}^{4}}
\end{equation}

\section{Conclusion and discussion}
Teleparallel gravity as the gauge theory for the translation group,
offers a viable gravitational theory for macroscopic matter. There
exist one class of teleparallel Lagrangians, called teleparallel
equivalent of general relativity (TEGR), which for all practical
purposes is empirically indistinguishable from general relativity
for scalar matter and electromagnetic fields. For finding possible
observational differences between TG and GR, we considered a 5
dimensional braneworld setup. The presence of the brane as a
boundary hypersurface embedded in the bulk, where all the ordinary
matter fields are confined to the brane and only gravitons can
propagate in the fifth dimension, offers an interesting opportunity
to study possible differences between TG and GR. In this paper,
using the results of ref \cite{BAN}, we constructed a RS-type
braneworld model in a teleparallel background. Starting from TEGR in
5 dimensions, we investigated possible local Lorentz invariance
violations in the effective 4 dimensional theory. In both cases of
Minkowski and FRW branes, the 4-D effective field equations found to
be local Lorentz invariant. Any possible difference between TG and
GR in the effective 4D dynamics, is a result of different junction
conditions in these two theories. In TG setup, the second junction
condition relates the jump in the superpotential tensor across the
brane to the matter content confined to the brane. This is in stark
difference to GR where the second junction condition involves the
extrinsic curvature. For the case of a FRW brane embedded in AdS
bulk, we studied both the background dynamics. FRW pentad
coefficients have been derived using the 5D field equations and
teleparallel junction conditions. The bulk field equations are
exactly the same as GR, however deriving the effective 4D equations
involves matching the discontinuities on both sides of the
teleparallel field equations via the junction conditions. As a
result of different junction conditions, the 4-D cosmological
evolution will be different in teleparallel gravity compared to GR.
For a quick illustration of practical results, we considered a
simple inflationary scenario where the 4D exponential expansion is
driven by a scalar field. In the slow-roll regime we found that the
slow-roll parameters are enhanced by braneworld modifications in
teleparallel gravity. This is quite different to the general
relativistic results in \cite{Mar} where the slow-roll parameters
were suppressed. This means that for a given potential, the
inflation will end sooner in teleparallel gravity than in general
relativity.

\end{document}